\documentclass[npg]{copernicus}

\titleheight{8cm}

\date{31 May 2010}

\begin{document}

\title{\textit{Preface}\\ ``Nonlinear processes in oceanic and atmospheric flows''}

\author[1]{A.~M.~Mancho}
\author[2]{S.~Wiggins}
\author[3]{A.~Turiel}
\author[4]{E.~Hern\'{a}ndez-Garc\'{\i}a}
\author[4]{C.~L\'{o}pez}
\author[3]{E.~Garc\'{\i}a-Ladona}

\affil[1]{Instituto de Ciencias Matem\'{a}ticas,
CSIC-UAM-UC3M-UCM, Serrano 121, 28006 Madrid, Spain}
\affil[2]{School of Mathematics, University of Bristol,
University Walk, Bristol BS8 1TW, UK} \affil[3]{Institut de
Ci\`encies del Mar, CSIC, Passeig Mar\'{\i}tim de la
Barceloneta 37-49, 08003 Barcelona, Spain} \affil[4]{Instituto
de F\'{i}sica Interdisciplinar y Sistemas Complejos IFISC
(CSIC-UIB), Campus Universitat de les Illes Balears,\newline
07122 Palma de Mallorca, Spain}
\affil[ ]{}
\affil[ ]{Nonlin. Processes Geophys. {\bf 17}, 283–-285 (2010) \\
www.nonlin-processes-geophys.net/17/283/2010/ \\
doi:10.5194/npg-17-283-2010 \\
Published under a Creative Commons Attribution 3.0 License}

\runningtitle{Nonlinear processes in oceanic and atmospheric flows}

\runningauthor{A.~M.~Mancho et al.}

\correspondence{A.~M.~Mancho \\(a.m.mancho@icmat.es)}

\firstpage{1}

\maketitle

\begin{abstract}
Nonlinear phenomena are essential ingredients in many oceanic and atmospheric
processes, and successful understanding of them benefits from
multidisciplinary collaboration between oceanographers, meteorologists,
physicists and mathematicians. The present Special Issue on ``Nonlinear
Processes in Oceanic and Atmospheric Flows'' contains selected contributions
from attendants to the workshop which, in the above spirit, was held in
Castro Urdiales, Spain, in July 2008. Here we summarize the Special Issue
contributions, which include papers on the characterization of ocean
transport in the Lagrangian and in the Eulerian frameworks, generation and
variability of jets and waves, interactions of fluid flow with plankton
dynamics or heavy drops, scaling in meteorological fields, and statistical
properties of El Ni\~no Southern Oscillation.
\end{abstract}

\section*{}

In recent years atmospheric and oceanic data sets arising from new
observational and computational capabilities have become widely available.
These data sets, and the variety of geophysical nonlinear phenomena that they
reveal, are giving rise to new challenges and opportunities that are
benefiting greatly from a multidisciplinary approach. Methods from diverse
areas of mathematics such as dynamical systems theory and statistics have
been combined with sophisticated computational methods and have been brought
to bear on a variety of data sets taken in very diverse physical settings.
These new approaches have been developed in collaborations between
mathematicians, physicists, oceanographers, and meteorologists.

\newpage

On 2--4~July 2008 such a group gathered for a workshop entitled ``Nonlinear
Processes in Oceanic and Atmospheric Flows'' held at Castro Urdiales,
Cantabria, Spain, with the generous support of the United States Office of
Naval Research (ONR Global), Consejo Superior de Investigaciones
Cient\'{\i}ficas (CSIC), Ministerio de Educaci\'{o}n y Ciencia (MEC), Centro
Internacional de Encuentros Matem\'{a}ticos (CIEM), Consolider i-MATH and
SIMUMAT. The fourteen papers in this Special Issue describe the breadth of
ideas and results discussed at this workshop and illustrate the exciting
opportunities for multidisciplinary collaborations in the oceanic and
atmospheric sciences.

The paper of \cite{Boucharel2009} introduces novel ideas from statistics,
taken from the field of financial mathematics, to perform a more detailed
diagnosis of the properties of the El Nino Southern Oscillation (ENSO). The
authors analyze data from the Zebiak-Cane model, models used by the
Intergovernmental Panel for Climate Change (IPCC), as well as in situ data.
Their analysis raises a number of provocative points and conclusions that
should be considered in the context of the fidelity of climate models in
general.

\citet{Rossi2009} analyze the interaction between eddy induced mixing and
phytoplankton distributions on small scale (1--100\,km) processes using
satellite data. Their analysis leads to the surprising conclusion that strong
mixing in nutrient-rich waters along Eastern Boundary Upwelling Systems
(e.g.\ the Benguela and Canary currents in the Atlantic Ocean, and the
Humboldt and California currents in the Pacific) appear to reduce, rather
than stimulate, growth of phytoplankton.

The paper by \citet{SanchezGarrido2009} addresses the behaviour of internal
solitary waves in a rotating and laterally confined domain, with emphasis in
the non-linear regime. According to the classical weakly non-linear theory,
energy is damped through radiation of secondary Poincar\'e gravity waves due
to rotational dispersion. \mbox{However}, under strong non-linearity
conditions, the energy damping is partially suppressed due to non-linear
wave-wave interactions. This leads to a regime where internal solitary waves
evolve into a slowly decaying packet of Kelvin waves that may propagate for a
long time. An understanding of phenomena of this type is fundamental for
obtaining a deeper insight into energy pathways in the oceans.

Using a variety of meteorological variables (derived either directly from
numerical simulations or from re-analysis which combine observed values with
numerical models assimilating them), \citet{Stolle2009} demonstrate that the
scaling properties of these variables can be explained in terms of underlying
multifractal cascades, beyond the usual, single-exponent characterization.
Their findings can be applied to improve the parametrization of numerical
models, as well as to validate the correctness of the implementation of
non-linear effects.

Using a simple idealized plankton model, \citet{McKiver2009} analyze the
importance of horizontal advection on phytoplankton biomass. They use a
single species model with multiple steady states depending on the values of
the carrying capacity, and show that small changes in the ratio of biological
to hydrodynamic time scales can greatly modify plankton production. As a
consequence, they argue that this effect may be a possible mechanism for
explaining plankton blooms or regime shifts in some oceanic regions.

\citet{Dellnitz2009} consider the fundamental issue of detecting regions in
the ocean that are coherent over an extended period of time. These
structures, such as gyres, are important with respect to the movement of heat
around the planet, distribution of nutrients, etc. The authors use a
realistic numerical model to study a 3-D coherent structure in the Southern
Ocean using a methodology based on transfer operators. They show that
transfer operators are a useful tool for identifying circulating pathways
across these structures.

\citet{Pierini2009} review the proposed ways to understand the bimodal
characteristics of the low-frequency variability of the Kuroshio System: a
state with the presence of a zonally elongated energetic meandering jet
alternating, on decadal time scales, with a state of a weaker jet with
reduced zonal penetration. The origin of such bimodality can be either in the
ocean response to changes of wind stress fields, and then due basically to
the atmospheric forcing of the ocean, or identified as intrinsic ocean
variability. As expected both aspects should be taken into account, but what
is remarkable is that the non-linear behavior of the bimodal system is quite
well reproduced and understood both quantitatively and qualitatively just by
considering the internal variability caused from homoclinic transitions
involving multiple equilibrium states of an ocean reduced gravity model under
steady wind forcing.

\citet{Zahnow2009} consider the effects of collision, coagulation and
fragmentation processes on the size distribution of heavy drops moving in a
turbulent fluid. The problem is relevant, for example, to the growth of cloud
droplets. The particle-based approach goes beyond simple transport models of
inertial particles, without the complications of a fully hydrodynamic
simulation. Scaling laws of mean sizes and distributions with respect to the
different flow and particle parameters are obtained by a combination of
numerical and theoretical arguments.

\citet{BranickiWiggins2010} give a critical analysis of the use of hyperbolic
trajectories, their stable and unstable manifolds, and finite time Lyapunov
exponents for revealing flow barriers and organized structures in
aperiodically time-dependent flows that exist only for a finite time. This is
a rapidly developing area due to the explosion in the availability of
observational and computational data sets for geophysical flows. This paper
takes a different point of view and describes a series of specific examples
that highlight different phenomena and their interpretation, as well as
problems and pathologies that can arise. Consequently, this paper provides
``benchmarks'' for the necessary further development of the theory and for
the application of these methods to complex geophysical flows.

\citet{Koszalka2010} explore how vertical transport within wind-forced eddies
is affected by stratification. They show that the wind energy injected at the
surface is transferred to depth through two stratification-dependent
mechanisms: vortex Rossby waves and near-inertial internal oscillations. In
view of their results on the role of wind-forced mesoscale vortices in the
transmission of wind energy into the ocean and vertical transport, the
authors stress the need to resolve the vertical transport and mixing by
mesoscale eddies in models designed to study oceanic circulation under
different climatological conditions.

\citet{Marie2010} studies mechanisms for the generation of zonal jets by
$\beta$-plane turbulence. The work begins with a simple situation -- a study
of linear perturbations of Rossby waves by zonal flow in an infinite
$\beta$-plane. He then considers a more realistic situation consisting of a
reduced-gravity model in a quasi-geostrophic setting and shows that
essentially the same results hold. This work provides insight into a complex
phenomenon resulting from a turbulence-mediated, subtle interaction, between
two very different scales.

The paper by \citet{Mendoza2010} applies a combination of Lagrangian tools,
some of them new and others well established, for studying transport in
velocity data sets obtained from altimetry over the Kuroshio current region.
The study shows how distinguished hyperbolic trajectories and their stable
and unstable manifolds can be computed in realistic data sets. It also
addresses how to achieve an accurate analysis of transport from the stable
and unstable manifolds. The method successfully characterizes the turnstile
mechanism across this area and this mechanism is shown to persist over the
spring months of year~2003.

\citet{BranickiMalekMadani2010} consider transport in a realistic
time-dependent-velocity data set obtained from a shallow water model of the
Chesapeake Bay. In this context they assess the limit of validity of 2-D
Lagrangian tools for analyzing estuarine flows. The 2-D Lagrangian analysis
of the surface flow captures the spatio-temporal variability of the
freshwater outflow events. The computation of finite time Lyapunov exponents
reveals a network of ridges, but these are often too short for a meaningful
transport analysis, while computation of stable and unstable manifolds of
relevant hyperbolic trajectories has the comparable challenge of first
computing the hyperbolic trajectories on a sufficiently long time interval.
It is anticipated that a symbiotic combination of these Lagrangian
diagnostics might overcome these difficulties. Their work points out that
further development of 3-D Lagrangian techniques is still required for
reliable transport analysis of complex coastal flows.

Hydrodynamic forcing is known to play an important role in plankton dynamics.
\citet{PerezMunuzuri2010} consider the influence of the spatial and temporal
scales of the flow on the spatial extension of a plankton bloom using a
reaction-diffusion-advection equation in which the reaction part models a
Nutrient-Phytoplankton-Zooplankton biological dynamics. Their analysis shows
that the bloom size is larger for certain length and time scales of the flow.
This is related to the fact that the balance of two processes, trapping fluid
inside eddies on the one hand, and mixing and diluting on the other hand, is
optimal for bloom growth at these particular length and time scales.

\begin{acknowledgements}

The workshop held at Castro Urdiales was possible thanks to the commitment of
its Organizing Committee: C.~L\'opez, A.~M.~Mancho, A.~Turiel,
E.~Garc\'ia-Ladona, E.~Hern\'andez-Garc\'ia, J.~A.~Jim\'enez-Madrid. Also
thanks to Ismael~Hern\'andez-Carrasco and Oriol~Pont for their assistance
during the event. The warm hospitality and support of the Cultural Centre
``La Residencia'' is also acknowledged.

The organization of the workshop was possible thanks to support from grants:
ONR Global (N00014-08-1-1035), CSIC Oceantech (PIF-0059-2006), Consolider
i-MATH C3-0103, CIEM, SIMUMAT S-0505-ESP-0158, MEC FIS2007-30844-E, CSIC
MP-38-AR.

\end{acknowledgements}

\end{document}